\documentclass[usenatbib]{mn2e}
\usepackage{times}
\usepackage{graphicx}

\pubyear{2006}

\author[Bagla and Prasad]
{J. S. Bagla$^1$ and Jayanti Prasad$^2$ \\
  $^1$, $^2$ Harish-Chandra Research Institute,  Chhatnag Road, Jhusi,
  Allahabad 211019, India. \\
  E-mail: $^1$ jasjeet@hri.res.in, $^2$ jayanti@hri.res.in}

\title[Effects of the size of cosmological N-Body simulations
  on physical quantities]
{Effects of the size of cosmological N-Body simulations
  on physical quantities --- I: Mass Function}

\label{firstpage}

\def\LaTeX{L\kern-.36em\raise.3ex\hbox{a}\kern-.15em
    T\kern-.1667em\lower.7ex\hbox{E}\kern-.125emX}

\def\lbx{L_{\rm box}}

\pagerange{\pageref{firstpage}--\pageref{lastpage}}

\begin{document}

\maketitle


\begin{abstract}
N-Body simulations are a very important tool in the study of formation of
large scale structures.
Much of the progress in understanding the physics of galaxy formation
and comparison with observations would not have been possible without N-Body
simulations.
Given the importance of this tool, it is essential to understand its
limitations as ignoring these can easily lead to interesting but
unreliable results.
In this paper we study the limitations due to the finite size
of the simulation volume. 
We explicitly construct the correction term arising due to a finite box size
and study its generic features for clustering of matter and also on mass
functions.  
We show that the correction to mass function is maximum near the scale of
non-linearity, as a corollary we show that the correction to the number
density of haloes of a given mass changes sign at this scale; the 
number of haloes at small masses is over estimated in simulations. 
This over estimate results from a delay in mergers that lead to formation of
more massive haloes. 
The same technique can be used to study corrections to other physical
quantities. 
The corrections are typically small if the scale of non-linearity is
much smaller than the box-size.  
However, there are some cases of physical interest in which the relative
correction term is of order unity even though a simulation box much larger
than the scale of non-linearity is used.  
Within the context of the concordance model, our analysis suggests that it is
very difficult for present day simulations to resolve mass scales smaller
than $10^2$~M$_\odot$ accurately and the level of difficulty increases as we
go to even smaller masses, though this constraint does not apply to
multi-scale simulations.
\end{abstract}


\begin{keywords}
methods: N-Body simulations, numerical -- gravitation -- cosmology : theory,
dark matter, large scale structure of the universe
\end{keywords}


\section{Introduction}

Large scale structures like galaxies and clusters of galaxies are
believed to have formed by gravitational amplification of small
perturbations.
For an overview and original references, see, e.g.,
\citet{1980lssu.book.....P,1999coph.book.....P,2002tagc.book.....P,2002PhR...367....1B}.  
Initial density perturbations were present at all scales that have been
observed \citep{2003ApJS..148..175S,2003MNRAS.346...78H,2004ApJ...607..655P}. 
Understanding evolution of density perturbations for such initial conditions
is essential for the study of formation of galaxies and large scale
structures.  
The equations that describe the evolution of density perturbations in an
expanding universe have been known for a long time \citep{1974A&A....32..391P}
and these are easy to solve when the amplitude of perturbations is
small.   
These equations describe the evolution of density contrast defined as
$\delta(\mathbf r, t) = (\rho(\mathbf r, t) - \bar\rho(t))/\bar\rho(t)$. 
Here $\rho(\mathbf r, t)$ is the density at point $\mathbf r$ and time $t$, and
$\bar\rho$ is the average density in the universe at time $t$.  
These are densities of non-relativistic matter, the component that clusters
at all scales and is believed to drive the formation of large scale structures
in the universe. 
Once density contrast at relevant scales becomes large, i.e., $|\delta| \geq
1$, the perturbation becomes non-linear and coupling with perturbations at
other scales cannot be ignored. 
The equation for evolution of density perturbations cannot be solved for
generic perturbations in this regime.
N-Body simulations
\citep{1998ARA&A..36..599B,1997Prama..49..161B,2005CSci...88.1088B} are often 
used to study the evolution in this regime. 
Alternative approaches can be used if one requires only a limited
amount of information and in such a case either quasi-linear approximation
schemes 
\citep{1970A&A.....5...84Z,1989MNRAS.236..385G,1992MNRAS.259..437M,1993ApJ...418..570B,1994MNRAS.266..227B,1995PhR...262....1S,1996ApJ...471....1H,2002PhR...367....1B}
or scaling relations
\citep{1977ApJS...34..425D,1991ApJ...374L...1H,1995MNRAS.276L..25J,2000ApJ...531...17Ka,1998ApJ...508L...5M,1994MNRAS.271..976N,1996ApJ...466..604P,1994MNRAS.267.1020P,1996MNRAS.278L..29P,1996MNRAS.280L..19P,2003MNRAS.341.1311S}
suffice. 

In cosmological N-Body simulations, we simulate a representative region of the
universe. 
This is a large but finite volume and periodic boundary conditions
are often used.   
Almost always, the simulation volume is taken to be a cube.
Effect of perturbations at scales smaller than the mass resolution of
the simulation, and of perturbations at scales larger than the box is
ignored.
Indeed, even perturbations at scales comparable to the box are under
sampled.

It has been shown that perturbations at small scales do not influence collapse
of perturbations at much larger scales in a significant manner
\citep{1974A&A....32..391P,1985ApJ...297..350P,1991MNRAS.253..295L,1997MNRAS.286.1023B,1998ApJ...497..499C}
if we study the evolution of the correlation function or power spectrum at
large scales due to gravitational clustering in an expanding universe. 
This is certainly true if the scales of interest are in the non-linear regime
\citep{1997MNRAS.286.1023B}. 
Therefore we may assume that ignoring perturbations at scales much smaller
than the scales of interest does not affect results of N-Body simulations. 
However, there may be other effects that are not completely understood
at the quantitative level \citep{2004astro.ph..8429B} even though these have
been seen only in somewhat artificial situations.

Perturbations at scales larger than the simulation volume can affect the
results of N-Body simulations. 
Use of periodic boundary conditions implies that the average density in the
simulation box is same as the average density in the universe, in other words
we ignore perturbations at the scale of the simulation volume (and at larger
scales).  
Therefore the size of the simulation volume should be chosen so that the
amplitude of fluctuations at the box scale (and at larger scales) is
ignorable.  
If the amplitude of perturbations at larger scales is not ignorable then
clearly the simulation will not be a faithful representation of the model
being studied. 
It is not obvious as to when fluctuations at larger scales can be considered
ignorable, indeed the answer to this question depends on the physical quantity
of interest, the model being studied and the specific length/mass scale of
interest as well.  

The effect of a finite box size has been studied using N-Body simulations and
the conclusions in this regard may be summarised as follows.  
\begin{itemize}
\item
If the amplitude of density perturbations around the box scale is small
($\delta < 1$) but not much smaller than unity, simulations underestimate
the correlation function though the number density of small mass haloes does
not change by much \citep{1994ApJ...436..467G,1994ApJ...436..491G}. 
In other words, the formation of small haloes is not disturbed but their
distribution is affected by non-inclusion of long wave modes.
\item
In the same situation, the number density of massive haloes drops
significantly
\citep{1994ApJ...436..467G,1994ApJ...436..491G,2005MNRAS.358.1076B}. 
\item
Effects of a finite box size modify values of physical quantities like the
correlation function even at scales much smaller than the simulation volume
\citep{2005MNRAS.358.1076B}.
\item
The void spectrum is also affected by finite size of the simulation volume if
perturbations at large scales are not ignorable \citep{1992ApJ...393..415K}.
\item
It has been shown that properties of a given halo can change significantly as
the contribution of perturbations at large scales is removed to the initial
conditions but the distribution of most internal properties remain unchanged
\citep{2005astro.ph.12281P}.  
\end{itemize}

In some cases, one may be able to devise a method to ``correct'' for the
effects of a finite box-size \citep{1994A&A...281..301C}, but such methods
cannot be generalised to all statistical measures or physical quantities. 

The effects of perturbations at scales larger than the box size can be added
using MAP (Mode Adding Procedure) after a simulation has been run
\citep{1996ApJ...472...14T}. 
This method makes use of the fact that if the box size is chosen to be large
enough then the contribution of larger scales can be incorporated by adding
displacements due to the larger scales independently of the evolution of the
system in an N-Body simulation. 
The motivation for development of such a tool is to enhance the range of
scales over which results of an N-Body simulation can be used by improving the
description at scales comparable to the box size. 
Such an approach ignores the coupling of large scale modes with small scale
modes and this again brings up the issue of what is a large enough scale for a
given model such that the effects of mode coupling can be ignored. 
Large scales contribute to displacements and velocities, and variations in
density due to these scales modify the rate of growth for small scales
perturbations \citep{1997MNRAS.286...38C}.

Effects of a finite box size modify values of physical quantities even at
scales much smaller than the simulation volume \citep{2005MNRAS.358.1076B}
(BR05, hereafter).  
In BR05, we suggested use of the fraction of mass in collapsed
haloes as an indicator of the effect of a finite box size.  
We found that if the simulation volume is not large enough, the fraction of
mass in collapsed haloes is underestimated.
As the collapsed fraction is less sensitive to box-size as compared to
measures of clustering, several other statistical indicators of clustering to
deviate significantly from expected values in such simulations.   
A workaround for this problem was suggested in the form of an ensemble of
simulations to take the effect of convergence due to long wave modes into
account \citep{2005ApJ...634..728S}, the effects of shear due to long wave
modes are ignored here.
It has also been shown that the distribution of most internal properties of
haloes, e.g., concentration, triaxiality and angular momentum do not change
considerably with the box size even though properties of a given halo may
change by a significant amount \citep{2005astro.ph.12281P}. 

There is a clear need to develop a formalism for estimating the effect of
perturbations at large scales on a variety of physical quantities.
Without such a formalism, we cannot decide in an objective manner whether a
simulation box size is sufficiently large or not.
In this paper we generalise the approach suggested in BR05 and write down an
explicit correction term for a number of statistical indicators of
clustering. 
This approach allows us to study generic properties of the expected correction
term in any given case, apart of course from allowing us to evaluate the
magnitude of the correction as compared to the expected value of the given
statistical indicator. 
We apply this technique to mass functions in this paper. 

\section{Basic Equations}

Initial conditions for N-Body simulations are often taken to be a realisation
of a Gaussian random field with a given power spectrum, for details see, e.g.,
\citet{1997Prama..49..161B,1998ARA&A..36..599B,2005CSci...88.1088B}.  
The power spectrum is sampled at discrete points in the $\mathbf k$ space
between the scales corresponding to the box size (fundamental mode) and the
grid size (Nyquist frequency/mode).  
Here $\mathbf k$ is the wave vector.  
Sampling of the power spectrum in initial conditions of N-Body simulations is
dense towards the Nyquist mode, but is sparse for modes near the fundamental
mode.   
Power spectra for density, potential and the velocity field are related to
each other in the linear regime\footnote{Density and Potential are related
  through the Poisson equation and hence the knowledge of power spectrum of
  one can be used to compute the power spectrum for the other quantity.
  These quantities at late times are obtained through evolution in which
  mode-coupling is significant and hence the effects of missing modes are not
  easy to quantify.  The sampling of initial conditions is, in our considered
  view, more relevant and easier to quantify than the effects of missing
  mode-coupling terms.  Therefore, in our discussion, we deal with the initial
  or the linearly evolved power spectra of various quantities.}.
The power spectra can be used to compute the second moment; either two point
functions or {\it rms} fluctuations. 
In view of the sampling of the power spectrum in initial conditions, the
second moment can be expressed as a sum over power spectrum at these points,
weighted by an appropriate window function.   

In the peak picture, most quantities of interest can be related to the two
point correlation function \citep{1986ApJ...304...15B}, therefore a method for
estimating box-size correction to the second moment can be used as a base for
computing correction for other physical quantities.   

\subsection{Clustering Amplitude}

We now present our approach for estimating the effects of a finite box size on
physical quantities in the linear limit.
We will illustrate our approach using {\it rms} fluctuations in mass
$\sigma(r)$, but as shown below, the basic approach can be generalised to any
other quantity in a straightforward manner.  
In general, $\sigma(r)$ may be defined as follows:
\begin{equation}
\sigma^2(r) = \int\limits_0^\infty \frac{dk}{k} \frac{k^3 P(k)}{2 \pi^2}
W^2(kr)
\end{equation}
Here $P(k)$ is the power spectrum of density contrast, $r$ is the comoving
length scale at which {\it rms} fluctuations are defined, $k = \sqrt{k_x^2 +
  k_y^2 + k_z^2}$ is the wave number and $W(kr)$ is the Fourier transform of
the window function used for sampling the density field.  
The window function may be a Gaussian or a step function in real or
$k$-space.
We choose to work with a step function in real space where $W(kr) = 9
\left(\sin{kr} - kr \cos{kr}\right)^2/(k^6 r^6)$, see e.g., \S{5.4} of
\citet{1993sfu..book.....P} for further details.    
In an N-Body simulation, the power spectrum is sampled only at specified
points in the $k$-space.  
In this case, we may write $\sigma^2(r)$ as a sum over these points. 
\begin{eqnarray}
\sigma^2(r,\lbx) &=& \frac{9}{V} \sum\limits_{\mathbf k}
P(k) \left[ \frac{\sin{kr} - kr
    \cos{kr}}{k^3 r^3}   \right]^2 \nonumber \\
&\simeq& \int\limits_{2\pi/\lbx}^{2\pi /{L_{\rm grid}}}
\frac{dk}{k} \frac{k^3 P(k)}{2 \pi^2} 9 \left[ \frac{\sin{kr} - kr
    \cos{kr}}{k^3 r^3}   \right]^2 \nonumber \\
&\simeq& \int\limits_{2\pi/\lbx}^\infty
\frac{dk}{k} \frac{k^3 P(k)}{2 \pi^2} 9 \left[ \frac{\sin{kr} - kr
    \cos{kr}}{k^3 r^3}   \right]^2 \nonumber \\
&=& \int\limits_0^\infty
\frac{dk}{k} \frac{k^3 P(k)}{2 \pi^2} 9 \left[ \frac{\sin{kr} - kr
    \cos{kr}}{k^3 r^3}   \right]^2 \nonumber \\
  && ~~~~ - 
\int\limits_0^{2\pi/\lbx}
\frac{dk}{k} \frac{k^3 P(k)}{2 \pi^2} 9 \left[ \frac{\sin{kr} - kr
    \cos{kr}}{k^3 r^3}   \right]^2 \nonumber \\
&=& \sigma_0^2(r) - \sigma_1^2(r,\lbx) 
\end{eqnarray}
Here $\sigma_0^2(r)$ is the
expected level of fluctuations in mass at scale $r$ for the given power
spectrum and $\sigma^2(r,\lbx)$ is what we get in an N-Body simulation at
early times. 
We have assumed that we can approximate the sum over the $k$ modes sampled in
initial conditions by an integral. 
Further, we make use of the fact that small scales do not influence large
scales to ignore the error contributed by the upper limit of the integral. 
This approximation is valid as long as the scales of interest are more than a
few grid lengths. 

\begin{figure}
\includegraphics[width=3.2in]{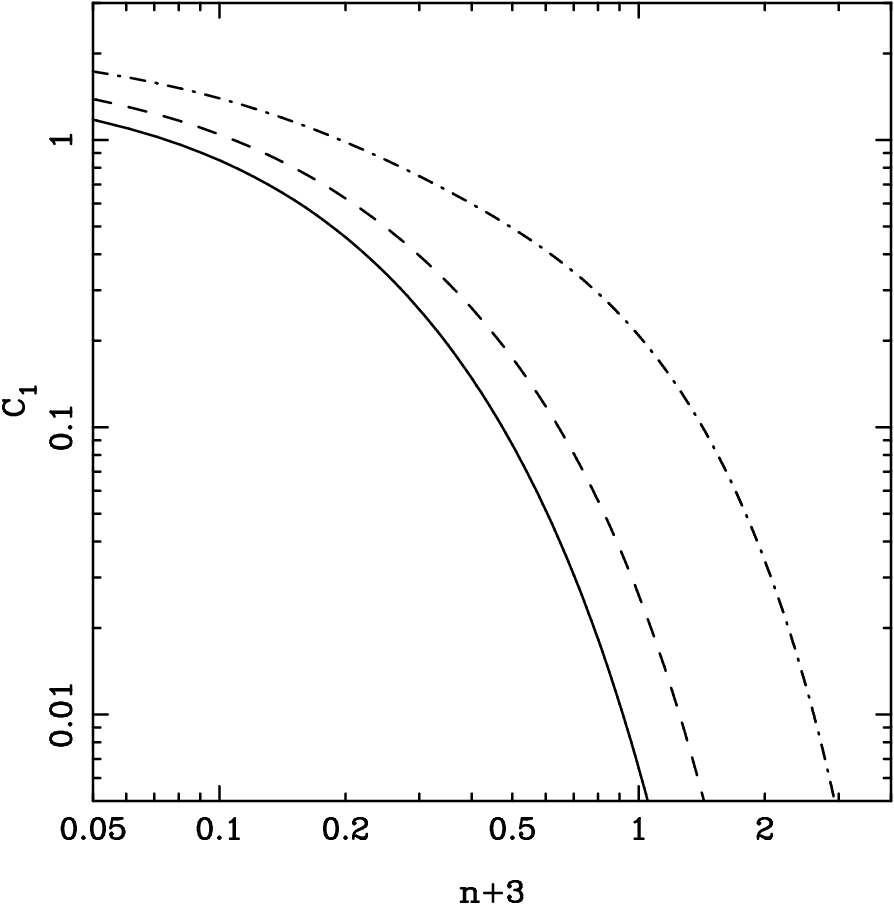}
\caption{This figure shows the first correction term $C_1$ (see
  Eqn.(\ref{eqn:c1})) for power law models with index $n$ normalised such that
  $\sigma_0^2(r_{nl})=1$.  The curves here are for $\lbx/r_{nl}=16$
  (dot-dashed curve), $\lbx/r_{nl}=128$ (dashed curve) and $\lbx/r_{nl}=512$
  (solid curve).  $C_1$ is plotted as a function of $n+3$ and we find that the 
  correction term increases as $n+3 \rightarrow 0$.  See text for more
  details.} 
\end{figure}

In the approach outlined above, the value of $\sigma^2$ at a given scale is
expressed as a combination of the expected value $\sigma_0^2$ and the
correction due to the finite box size $\sigma_1^2$.  
Here $\sigma_0^2$ is independent of the box size and depends only on the power
spectrum and the scale of interest.  
It is clear than $\sigma^2(r,\lbx) \leq \sigma_0^2(r)$ and also
$\sigma_1^2(r,\lbx) \geq 0$. 
It can also be shown that for hierarchical models, $d\sigma_1^2(r,\lbx)/dr
\leq 0$, i.e., $\sigma_1^2(r,\lbx)$ increases or saturates to a constant value
as we approach small $r$. 

If the scale of interest is much smaller than the box-size $\lbx$ then,
\begin{eqnarray}
\sigma_1^2(r,\lbx) &=& \int\limits_0^{2\pi/\lbx}
\frac{dk}{k} \frac{k^3 P(k)}{2 \pi^2} 9 \left[ \frac{\sin{kr} - kr
    \cos{kr}}{k^3 r^3}   \right]^2 \nonumber \\
&\simeq& \int\limits_0^{2\pi/\lbx} \frac{dk}{k} \frac{k^3 P(k)}{2 \pi^2} 
\nonumber \\
&& - 
\frac{r^2}{5} \int\limits_0^{2\pi/\lbx} \frac{dk}{k} \frac{k^5 P(k)}{2 \pi^2}
\nonumber \\
&& + \frac{3 r^4}{175} \int\limits_0^{2\pi/\lbx} \frac{dk}{k} \frac{k^7 P(k)}{2
  \pi^2} + \mathcal{O}(r^6) \\
&=& C_1 - C_2 r^2 + C_3 r^4 + \mathcal{O}(r^6) \label{eqn:sig1_series}
\end{eqnarray}
The small parameter in the expansion is $r/\lbx$.
This expansion is useful if $k^3P(k)$ goes to zero as we approach $k=0$. 
It is interesting to note that the first term is scale independent. 
The numerical values of $C_i$ can be used to estimate the scale below which
$\sigma_1$ can be approximated by a constant. 
Later terms are scale dependent and the noteworthy feature is that modes 
closer to $2\pi/\lbx$ contribute more significantly to the integral for most
models. 

It is noteworthy that the first term, $C_1$, has the same value for all
choices of window functions that approach unity at small $k$.  
By virtue of this fact, $C_1$ is also the correction for the two point
correlation function $\xi(r)$ at sufficiently small scales. 

\begin{figure}
\includegraphics[width=3.2in]{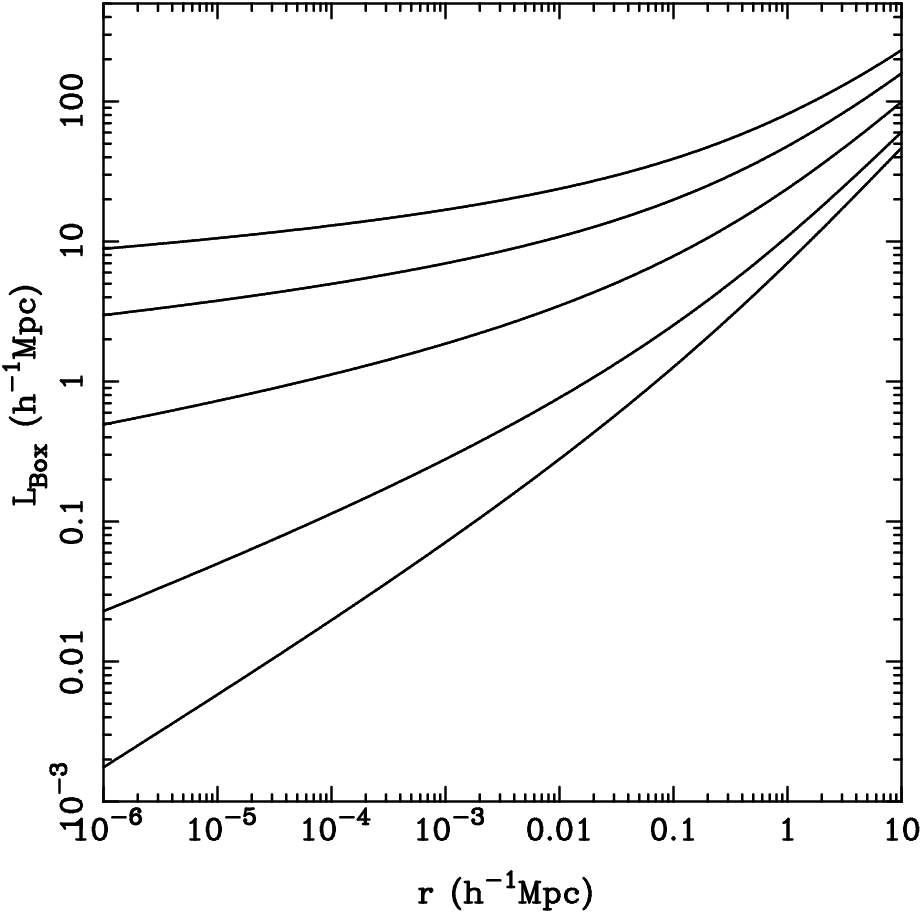}
\includegraphics[width=3.2in]{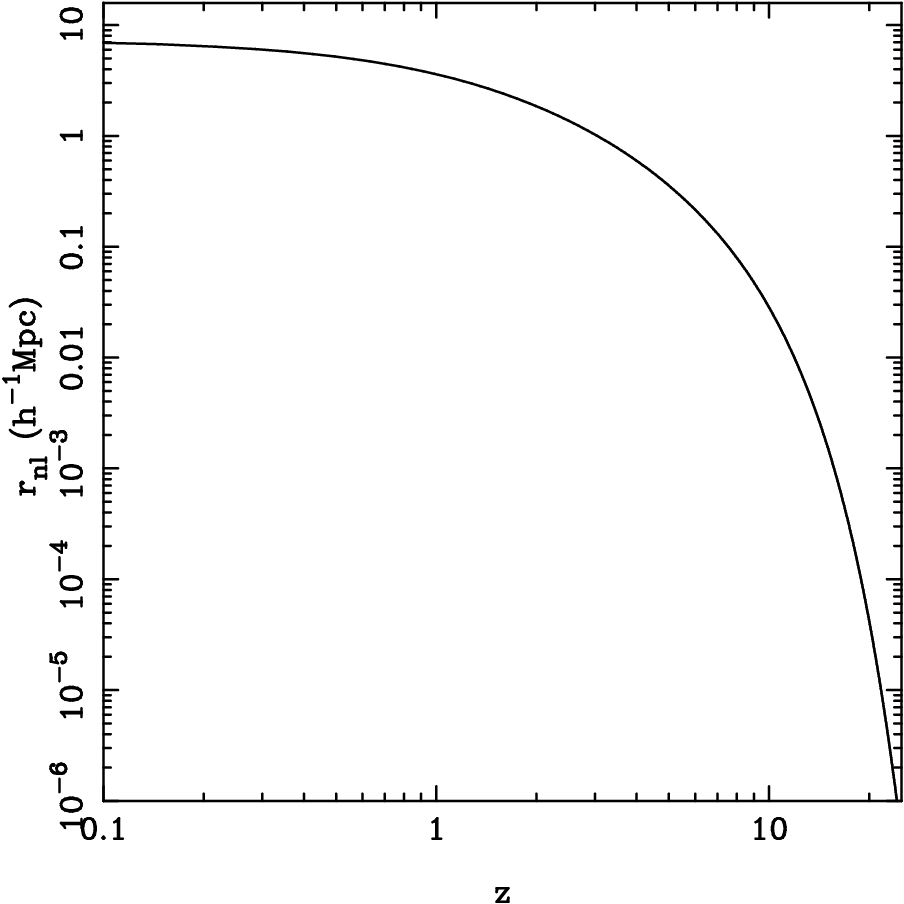}
\caption{The top panel shows curves of constant $C_1(\lbx)/\sigma_0^2(r)$ on
  the $r-\lbx$ plane for the $\Lambda$CDM model (see text for details).  Lines 
  mark $C_1/\sigma_0^2 = 0.01$, $0.03$, $0.1$, $0.3$ and $0.5$, from top to
  bottom.  The lower panel shows the scale of non-linearity $r_{nl}$ as a
  function of redshift for the $\Lambda$CDM model.}
\end{figure}

At large scales $\sigma_0^2(r)$ and $\sigma_1^2(r,\lbx)$ have a
similar magnitude and the {\it rms} fluctuations in the simulation become
negligible compared to the expected values in the model. 
As we approach small $r$ the correction term $\sigma_1^2(r,\lbx)$ is constant
and for most models it becomes insignificant in comparison with
$\sigma_0^2(r)$.  
In models where $\sigma_0^2(r)$ increases very slowly at small scales or
saturates to a constant value, the correction term $\sigma_1^2$ can be
significant at all scales.  
This can be seen from the expression for $C_1$ for power law
models ($P(k) = A k^n$)
\begin{equation}
C_1 = \frac{1}{n+3} ~~ \frac{A}{2 \pi^2} ~ \left(\frac{2
    \pi}{\lbx}\right)^{n+3} \label{eqn:c1}
\end{equation}
Clearly, this term becomes more and more significant as $n \rightarrow -3$. 
Figure~1 illustrates this point, here $C_1$ is shown as a function of $n+3$.   
We fix $A$ by choosing a scale of non-linearity $r_{nl}$ such that
$\sigma_0(r_{nl})=1$.  
Curves are plotted for three values of $\lbx/r_{nl}$: $\lbx/r_{nl}=16$
(dot-dashed curve), $\lbx/r_{nl}=128$ (dashed curve) and $\lbx/r_{nl}=512$
(solid curve).  
As $\sigma_0$ is unity at the scale of non-linearity and $C_1$ is the first
order correction, clearly we require $C_1 \ll 1$ for the error due to box size
to be small and hence ignorable.  
If we fix $C_1 \leq 0.1$ then we can simulate $n=-1$ with $\lbx/r_{nl}=16$ but
for more negative indices we require a larger separation between the box size
and the scale of non-linearity.  
We can just about manage $n=-2.3$ with $\lbx/r_{nl}=128$ with the same
threshold on error, and with $\lbx/r_{nl}=512$ we can go up to $n=-2.5$. 
As N-Body simulations are most useful for studying non-linear evolution, even
the largest simulations possible today are left with a small range of
scales over which $\sigma_0 \geq 1$ for $n \leq -2.0$. 
This shows the pitfalls of simulating models with $n \simeq -3$ over
the entire range of scales. 

Figure~2 (top panel) shows lines of constant $C_1/\sigma_0^2$ in the $\lbx-r$
plane for the $\Lambda$CDM.  
We chose $n=1$, $h=0.7$, $\Omega_\Lambda=0.7$, $\Omega_{nr}=0.3$ and
$\sigma_8=0.9$. 
We ignored the effects of Baryons on the power spectrum. 
From top to bottom, the lines correspond to $C_1/\sigma_0^2 = 0.01$, $0.03$,
$0.1$, $0.3$ and $0.5$.  
It is noteworthy that a box-size smaller than $0.5$Mpc is precluded if we
insist on $C_1(\lbx)/\sigma_0^2(r) \leq 0.1$, irrespective of the scale of
interest. 
This implies that we cannot expect to simulate scales smaller than about
$0.5$~kpc in the $\Lambda$CDM model without considerable improvement in the
dynamic range of cosmological N-Body simulations.  
As we are using linearly evolved quantities for our argument, the
comments on box size are valid irrespective of the redshift up to which the
simulation is run. 
The contours do not change if we use $\sigma_1^2$ instead of $C_1$.

The lower panel of the same figure shows the scale of non-linearity for the
$\Lambda$CDM model as a function of redshift.

This formalism can be used to estimate corrections for other estimators of
clustering.   
For reference, we have given expressions equivalent to
Eqn.(\ref{eqn:sig1_series}) for the correction to $\xi$ and $\bar\xi$ in
Table~1. 

\begin{table}
\caption{This table lists corrections due to a finite box-size to indicators
  of clustering in the limit $r \ll \lbx$.  These expressions are equivalent
  to Eqn.(\ref{eqn:sig1_series}) and constants $C_i$ are the same as in that
  equation.} 
\begin{center}
\begin{tabular}{||c|c||}
\hline
\hline
Indicator & Correction \\
\hline
\hline
$\xi(r)$ & $C_1 - \frac{5}{6} C_2 r^2 + \frac{35}{72} C_3 r^4 +
\mathcal{O}(r^6)$ \\
\hline
$\bar{\xi}(r)$ & $ C_1 - \frac{1}{2} C_2 r^2 + \frac{5}{24} C_3 r^4 +
\mathcal{O}(r^6)$ \\
\hline
\hline
\end{tabular}
\end{center}
\end{table}

\subsection{Velocities}

We can use the method outlined above to estimate finite box corrections to the
velocity field.  
Velocities and density contrast are related to one another
\citep{1980lssu.book.....P} in the linear regime.  
The power spectra for these two are related as $P_v(k) \propto P(k)/k^2$.  
Thus bulk velocities at any given scale get a larger contribution from the
power spectrum at large scales (small $k$) than density fluctuations.  
This implies that the correction term must be more significant for velocities
than the equivalent correction for the clustering amplitude.
We will discuss the corrections in velocity field in detail in a follow up
paper.

\subsection{Mass Function}

We can use the explicit correction for {\it rms} fluctuations ($\sigma$) to
estimate the correction for mass functions of haloes.  
We use the Press-Schechter approach
\citep{1974ApJ...187..425P,1991ApJ...379..440B}, but we also
give results for the Sheth-Tormen mass function
\citep{1999MNRAS.308..119S,2001MNRAS.323....1S} in order to demonstrate that
our results are generic in nature.  

The mass fraction in collapsed haloes with mass greater than $M$ 
is given in the Press-Schechter model by
\begin{eqnarray}
F(M,\lbx) &=& {\rm erfc}\left(\frac{\delta_c}{\sigma(M,\lbx)\sqrt{2}}\right) 
\nonumber \\
&=&
\frac{2}{\sqrt{\pi}} \int\limits_{\delta_c/\sigma(M,\lbx)\sqrt{2}}^\infty
  \exp\left[-x^2 \right] dx
\label{eqn:psmf}
\end{eqnarray}
Where $\delta_c$ ($\simeq 1.68$ for Einstein-de Sitter cosmology) is a
  parameter\footnote{In the spherical collapse model, this is the linearly
  extrapolated density contrast at which we expect the halo to virialise
  \citep{1972ApJ...176....1G}.} and $M$ is related to the scale $r$ through
the usual relation. 
We can write $F$ as the contribution expected in the limit $\lbx
\rightarrow \infty$ and a correction due to the finite box size.  
\begin{eqnarray}
F(M,\lbx) &=& \frac{2}{\sqrt{\pi}}
  \int\limits_{\delta_c/\sigma_0(M)\sqrt{2}}^\infty 
  \exp\left[-x^2 \right] dx 
\nonumber \\
&& - \frac{2}{\sqrt{\pi}}
  \int\limits_{\delta_c/\sigma_0(M)\sqrt{2}}^{\delta_c/\sigma(M,\lbx)\sqrt{2}} 
  \exp\left[-x^2 \right] dx \nonumber \\
&=& F_0(M) - F_1(M,\lbx)
\end{eqnarray}
The correction to $F(M)$ due to the finite box size always leads to an
under-estimate as $F_1(M,\lbx)$ is always positive. 
This is consistent with what we found in BR05.   
However, $F_1(M,\lbx)$ is not a monotonic function of $M$ as it goes to zero
at small as well as large $M$.  
At small $M$ ($M \ll M_{nl}$)\footnote{$M_{nl}$ is the mass corresponding to
  the scale where $\sigma_0 = 1$ and we shall assume that $\lbx$ is much
  larger than this scale.}, the limits of the integral differ by a very 
small amount.  
This difference ($\delta_c \sigma_1^2 / 2 \sqrt{2} \sigma_0^3$) keeps on
decreasing as we get to small $M$ while the integrand remains finite.  
Therefore we expect $F_1$ to decrease at small $M$.  
At these scales, we can write an approximate expression for $F_1(M)$:
\begin{equation}
F_1(M) \simeq \frac{\delta_c}{\sqrt{2\pi}} \frac{\sigma_1^2}{\sigma_0^3}
\exp\left[ - \frac{\delta_c^2}{2\sigma_0^2} \right] ~~~ .
\label{eqn:f1approx}
\end{equation}
This clearly decreases as we go to small $M$: $\sigma_1$ goes over to the
constant $C_1$ and $\sigma_0$ keeps increasing. 

\begin{figure}
\includegraphics[width=3.2in]{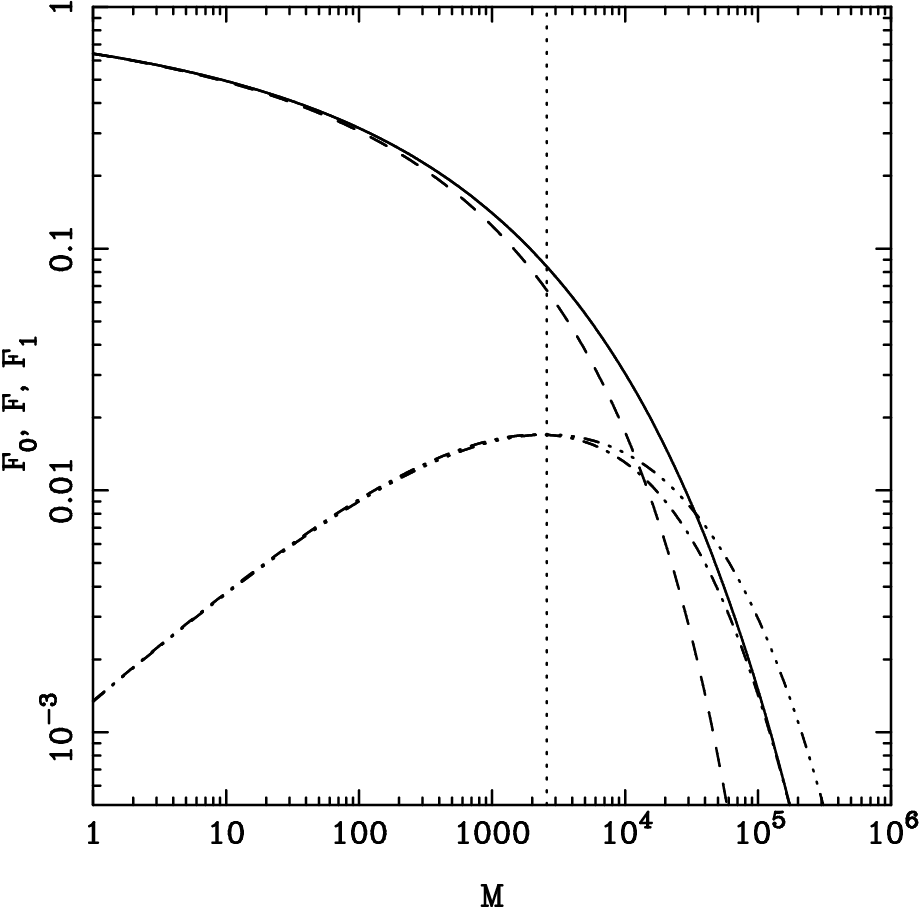}
\caption{The Press-Schechter mass function and correction terms are plotted as
a function of mass.   $F_0(M)$ (solid curve), $F(M)$ (dashed curve) and
$F_1(M)$ (dot-dashed curve) are shown here.  The scale where $\sigma_0 =
\delta_c/\sqrt{3}$ is marked with a vertical dotted line, we see that this
estimate coincides with the maximum of $F_1(M)$.  The correction term $F_1(M)$
is more than $10\%$ of $F_0(M)$ at this scale.  Also shown is the approximate
expression Eqn.(\ref{eqn:f1approx}) for $F_1(M)$ (dot-dot-dot-dashed curve)
and we note that it follows the actual curve to masses greater than $M_{nl}$.
Mass here is plotted in units of mass of each particle and we assumed that the
scale of non-linearity is $8$ grid lengths.} 
\end{figure}

\begin{figure}
\includegraphics[width=3.2in]{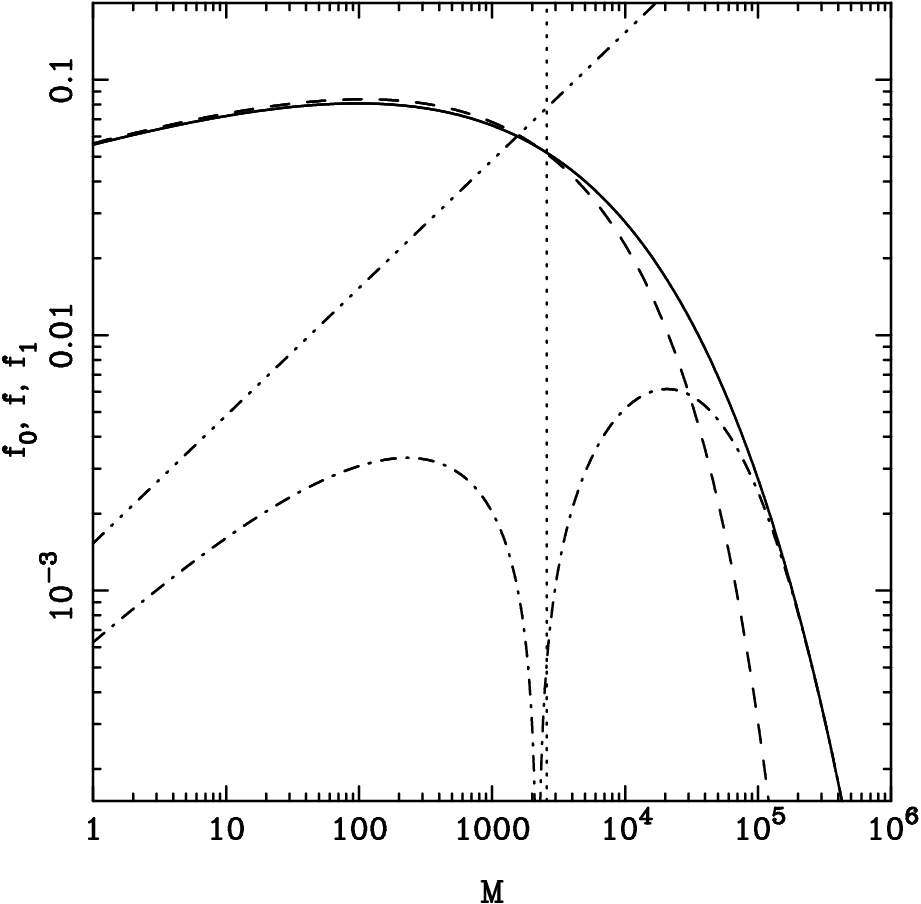}
\caption{The Press-Schechter multiplicity function and correction terms are
  plotted as a function of mass.   $f_0(M)$ (solid curve), $f(M)$ (dashed
  curve) and $f_1(M)$ (dot-dashed curve) are shown here.  The scale where
  $\sigma_0 = \delta_c/\sqrt{3}$ is marked with a vertical dotted line, we see
  that this estimate coincides with change of sign for $f_1(M)$.  At scales
  below this, the correction term $f_1(M)$ is positive and hence there are
  more haloes in simulation than expected in the model.  Also shown is the
  approximate expression for $f_1(M)$ (dot-dot-dot-dashed curve).  Mass here
  is plotted in units of mass of each particle and we assumed that the scale
  of non-linearity is $8$ grid lengths.}  
\end{figure}

At large $M$ ($M \gg M_{nl}$), both $\sigma(M,\lbx)$ and $\sigma_0(M)$ are
small and the limits of the integral cover the region where the integrand is
very small. 
Thus we expect $F_1(M,\lbx)$ to become smaller as we go to larger $M$ in this
regime.  
At these scales, we also expect $F_0$ and $F_1$ to become almost equal while
$F(M)$ goes to zero faster than either term. 
Therefore $F_1(M,\lbx)$ must have a maxima at an intermediate scale. 
The scale at which the maxima occurs can be found by solving the following
equation.  
\begin{eqnarray}
\frac{d\log\sigma_1^2}{d\log\sigma_0^2} &=& - \frac{
  \sigma_0^2}{\sigma_1^2} \left[\frac{\sigma}{\sigma_0}\left(1 -
  \frac{\sigma_1^2}{\sigma_0^2}\right)
  \exp\left[\frac{\delta_c^2\sigma_1^2}{2\sigma^2\sigma_0^2}\right]  
 - 1 \right] \nonumber \\
&\simeq & \frac{3}{2} - \frac{\delta_c^2}{2\sigma_0^2} 
\label{eqn:Fmax}
\end{eqnarray}
Here, the second equation is obtained if $\sigma_1 \ll \sigma_0$. 
If $\lbx \gg r_{nl}$, where $r_{nl}$ is the scale of non-linearity then
$\sigma_1$ is very well approximated by the Taylor series
Eqn.(\ref{eqn:sig1_series}) around this scale and $\sigma_1$ is a very slowly
varying function of scale.  
Thus $F_1(M,\lbx)$ has a maxima at $\sigma_0 = \delta_c^2/3 \sim 1$ if
the first term in Eqn.(\ref{eqn:sig1_series}) is a good approximation for
$\sigma_1$. 
If scale dependent terms in Eqn.(\ref{eqn:sig1_series}) are not ignorable then
the maxima of $F_1(M,\lbx)$ shifts to smaller scales (larger $\sigma_0$) in a
manner that depends on the power spectrum and box-size $\lbx$. 

Figure~3 shows the Press-Schechter mass function $F(M)$ for a power law
model with $n=-2$, $\lbx/r_{nl} = 16$.
We have plotted $F_0(M)$ (solid curve), $F(M)$ (dashed curve) and $F_1(M)$
(dot-dashed curve) as a function of $M$.  
The scale where $\sigma_0 = \delta_c/\sqrt{3}$ is marked with a vertical dotted
line, we see that this estimate is close to the maximum of $F_1(M)$.  
The correction term $F_1(M)$ is more than $10\%$ of $F_0(M)$ at this scale. 
Also shown is the approximate expression Eqn.(\ref{eqn:f1approx}) for $F_1(M)$
(dot-dot-dot-dashed curve) and we note that it follows the actual curve to
masses greater than $M_{nl}$.
This figure illustrates all the generic features of corrections to mass
function that we have discussed above.  

The multiplicity function $f$ is often defined as the fraction of mass in a
logarithmic interval in mass:
\begin{eqnarray}
f(M,\lbx) d\log{M} &=& - \frac{\partial F(M,\lbx)}{\partial \log{M}} d\log{M}
\nonumber \\ 
\Rightarrow f(M,\lbx) &=& - \frac{d F_0(M)}{d \log{M}} +
\frac{\partial F_1(M,\lbx)}{\partial \log{M}} \nonumber \\
&=& f_0(M) - f_1(M,\lbx) .
\end{eqnarray}
It is not possible to reduce this expression further while writing
the correction term due to the finite box size
separately.  
We can, however, ascertain generic properties of the correction term
$f_1(M,\lbx)$ from our understanding of $F_1(M,\lbx)$. 
At large $M$, $f_1$ is positive as $F_1(M,\lbx)$ decreases with increasing
$M$.  
Thus the mass fraction of haloes in this mass range is underestimated in
simulations. 
For typical models and simulations, this is the most significant effect of a
finite box size.  

We know that $f_1$ has a zero near the scale of non-linearity as $F_1$ has a
maxima here. 
Thus there is a scale where corrections for the multiplicity function due to a
finite box size vanish. 
At smaller scales, the slope of $F_1$ and hence $f_1$ changes sign and the
correction to mass fraction in haloes is positive. 
A finite box size leads to an {\it over estimate} of number of low mass
haloes.  
This over estimate is caused by absence of long wave modes, as the low mass
haloes do not merge to form the high mass haloes. 

The magnitude of over estimate depends on $\sigma_1$, and hence on the slope
of the power spectrum and $\lbx$.  
In the limit of $M \ll M_{nl}$, we can use Eqn.(\ref{eqn:f1approx}) to compute
the magnitude of over estimate: 
\begin{equation}
f(M) \simeq  f_0(M) + \frac{3 \delta_c}{\sqrt{2\pi}}
  \frac{C_1}{\sigma_0^4}  \left|   \frac{d\sigma_0}{d\log{M}} \right| ~~~ .
\end{equation} 
Here we have ignored the contribution of the exponential term in
Eqn.(\ref{eqn:f1approx}). 
The correction term scales as $M^{(n+3)/2}$ for power law models, thus it is
significant even at small mass scales if $n \simeq -3$. 
Clearly, the term is also large for CDM like power spectra if the slope
of the power spectrum is close to $-3$ at all scales in the simulation
volume. 

Figure~4 shows the Press-Schechter multiplicity function and correction terms
as a function of mass for the model used in Figure~3 (Power law model with
$n=-2$, $\lbx/r_{nl}=16$.).   
The expected multiplicity function $f_0(M)$ (solid curve), what is expected in
the simulation $f(M)$ (dashed curve) and the correction term $f_1(M)$
(dot-dashed curve) are shown here.  
The scale where $\sigma_0 = \delta_c/\sqrt{3}$ is marked with a vertical dotted
line, we see that this almost coincides with change of sign for
$f_1(M)$\footnote{The change of sign happens at $\sigma_0 = 1$
  instead of $\sigma_0=0.97$ drawn here with $\delta_c=1.68$.}.  
At scales below this, the correction term $f_1(M)$ is positive and hence there
are more haloes in the simulation than expected in the model.  
The relative magnitude of the correction term is large for $M > M_{nl}$ and
this is the most significant effect of a finite box-size on the mass
function. 
The over estimate of the multiplicity function is typically a sub-dominant
effect, as it is for the model shown here.  
However, as we shall see below, this effect can be very significant in some
situations. 
Also shown in the figure is the approximate expression for $f_1(M)$
(dot-dot-dot-dashed curve) in the limit $M \ll M_{nl}$. 
Unlike the approximation for $F_1(M)$ which is accurate over a large range of
scales, this is expected to be valid only in the limit of $M \ll M_{nl}$ and
indeed, is off by about a factor of two at the smallest scales shown here. 
However, it is a good approximation if we go to even smaller masses. 
We note that for this model, the over estimate of multiplicity function due to
the finite box is small and therefore is difficult to detect.  
For this model, $C_1/\sigma_0^2 \simeq 0.2$ at the scale of non-linearity and
is smaller than $0.1$ at scales where the over estimate in $f(M)$ is maximum. 
At small scales, $f_1/f_0$ is typically of the same order of magnitude as
$C_1/\sigma_0^2$. 

\begin{figure}
\includegraphics[width=3.2in]{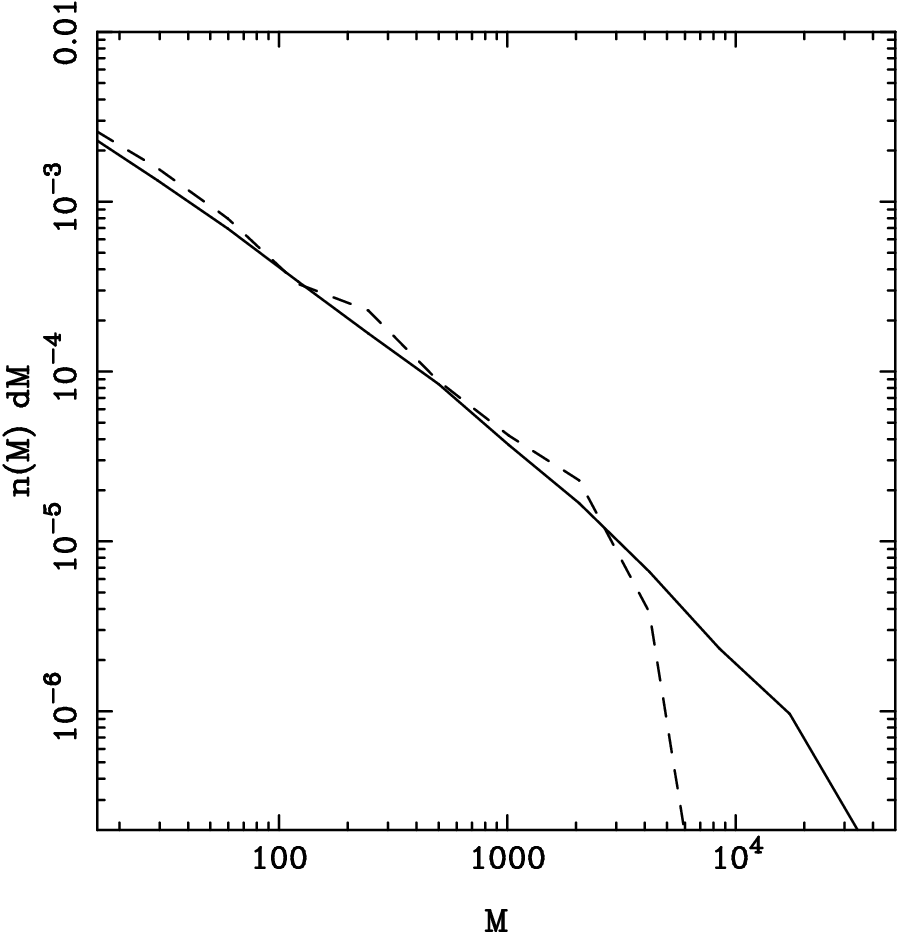}
\caption{Shown here is the number density of haloes $n(M) dM$ in the mass
  range $M~-~M+dM$ for these two simulations.  The solid line shows the number
  density of haloes in the reference simulation ($\lbx = 256$). Number
  density of haloes in the simulation with $\lbx = 64$Mpc is shown by the
  dashed line.}  
\end{figure}

\subsubsection{Sheth-Tormen Mass Function}

We now give corresponding formulae for the Sheth-Tormen mass function
\citep{1999MNRAS.308..119S,2001MNRAS.323....1S}. 
The definition of mass function (Eqn.\ref{eqn:psmf}) is modified to:
\begin{equation}
F(M,\lbx) = \frac{2}{\sqrt{\pi}}
 \int\limits_{\delta_c/\sigma(M,lbx)\sqrt{2}}^\infty 
 A(1+x^{-2q}) \exp\left[-x^2 \right] dx
\end{equation}
In the limit of $A=0.5$ and $q=0$ this is identical to the usual
Press-Schechter mass function (Eqn.\ref{eqn:psmf}).
The maxima of the correction term ($F_1(M,\lbx)$) occurs when the following
equation is satisfied:
\begin{eqnarray}
\frac{d\log\sigma_1^2}{d\log\sigma_0^2} &=& - \frac{
  \sigma_0^2}{\sigma_1^2} \left[\frac{\sigma}{\sigma_0} \left(1 -
  \frac{\sigma_1^2}{\sigma_0^2}\right) \right. \nonumber \\
&& ~~
 \left. 
  \exp\left[\frac{\delta_c^2\sigma_1^2}{2\sigma^2\sigma_0^2}\right]
{  \frac {1+  \left (\frac {\delta_c}{\sqrt {2} \sigma_0} \right )^{-2q} }
{1+  \left (\frac {\delta_c}{\sqrt {2} \sigma} \right )^{-2q} }}
 - 1 \right] \nonumber \\
&\simeq & \frac{3}{2} - \frac{\delta_c^2}{2\sigma_0^2} 
- q \left (\frac {\delta_c}{\sqrt {2} \sigma_0} \right )^{-2q}
\end{eqnarray}
As before, this reduces to the expression in the Press-Schechter case
(Eqn.\ref{eqn:Fmax}) in the limit $q=0$. 
The qualitative features of the finite box correction to mass function
are the same for the two prescriptions and may be considered to be generic.  
For reference, we write approximate expressions for correction to the mass
function $F(M)$:
\begin{equation}
F_1 \simeq \frac {\delta_c}{\sqrt {2 \pi}} \frac {\sigma_1^2} {\sigma_0^3} 
\exp\left[ -\frac{ \delta_c^2 }{2\sigma_0^2} \right]
A \left[ 1 + \left( \frac{\delta_c} {\sqrt{2} \sigma_0} \right)^{-2q} \right ]
\end{equation}
and the multiplicity function $f(M)$:
\begin{equation}
f_1 = \frac{3 \delta_c}{\sqrt{2 \pi}} \frac{C_1} {\sigma_0^4} \left(
  \frac{d\sigma_0}{d\log M} \right) A \left[ 1 + \left(1 - \frac{2q}{3}
  \right)  \left( \frac{\delta_c}{\sqrt{2} \sigma_0} \right)^{-2q} \right]  
\end{equation}
for the Sheth-Tormen mass function.

\begin{figure*}
\centering
\begin{tabular}{cc}
\begin{minipage}{3in}
\centering
\includegraphics[width=3in]{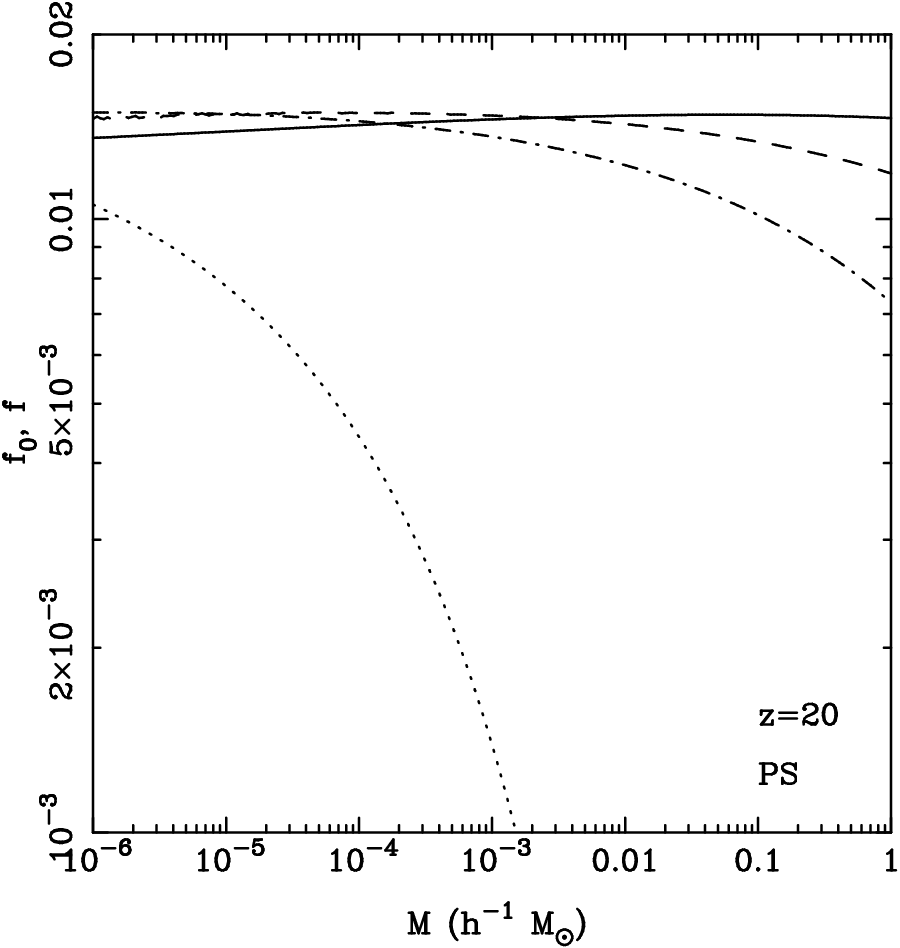}
\end{minipage}
&
\begin{minipage}{3in}
\centering
\includegraphics[width=3in]{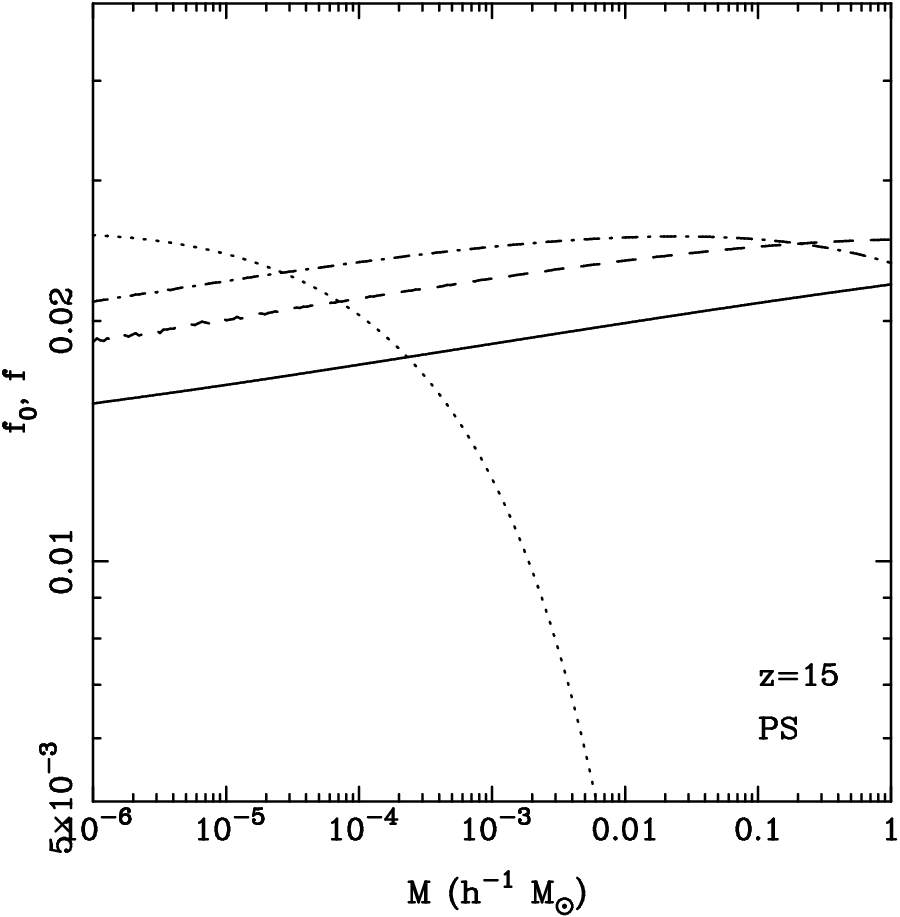}
\end{minipage}
\\
\begin{minipage}{3in}
\centering
\includegraphics[width=3in]{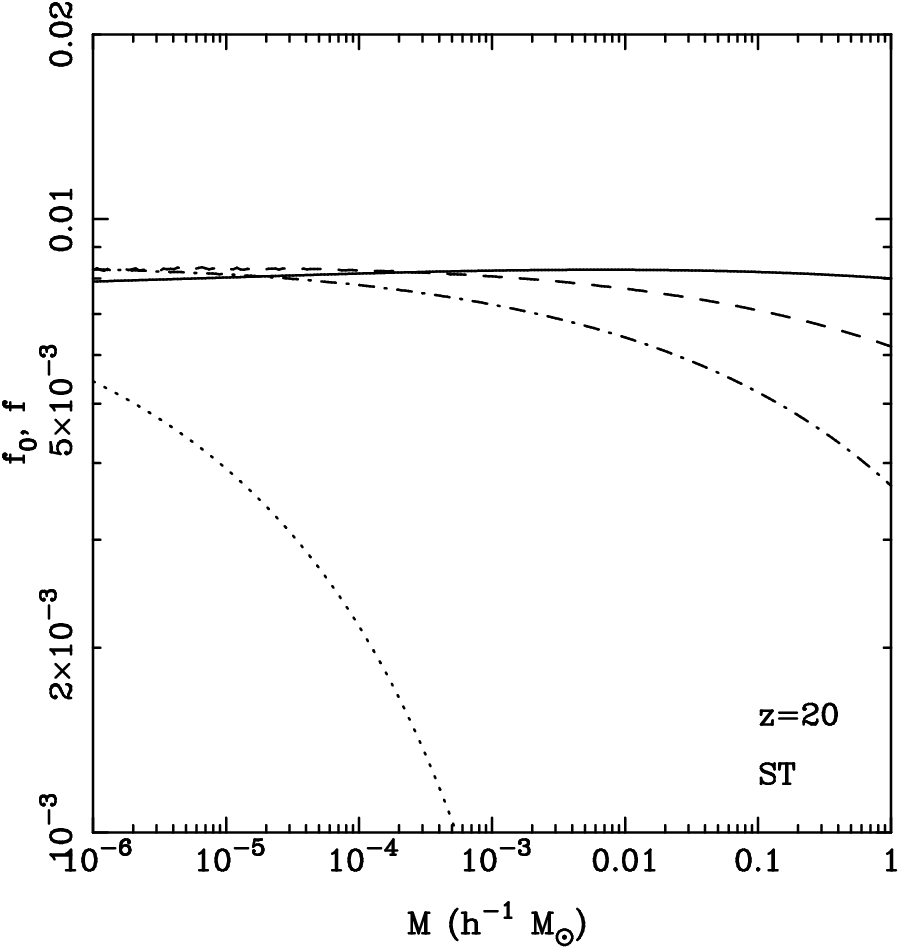}
\end{minipage}
&
\begin{minipage}{3in}
\centering
\includegraphics[width=3in]{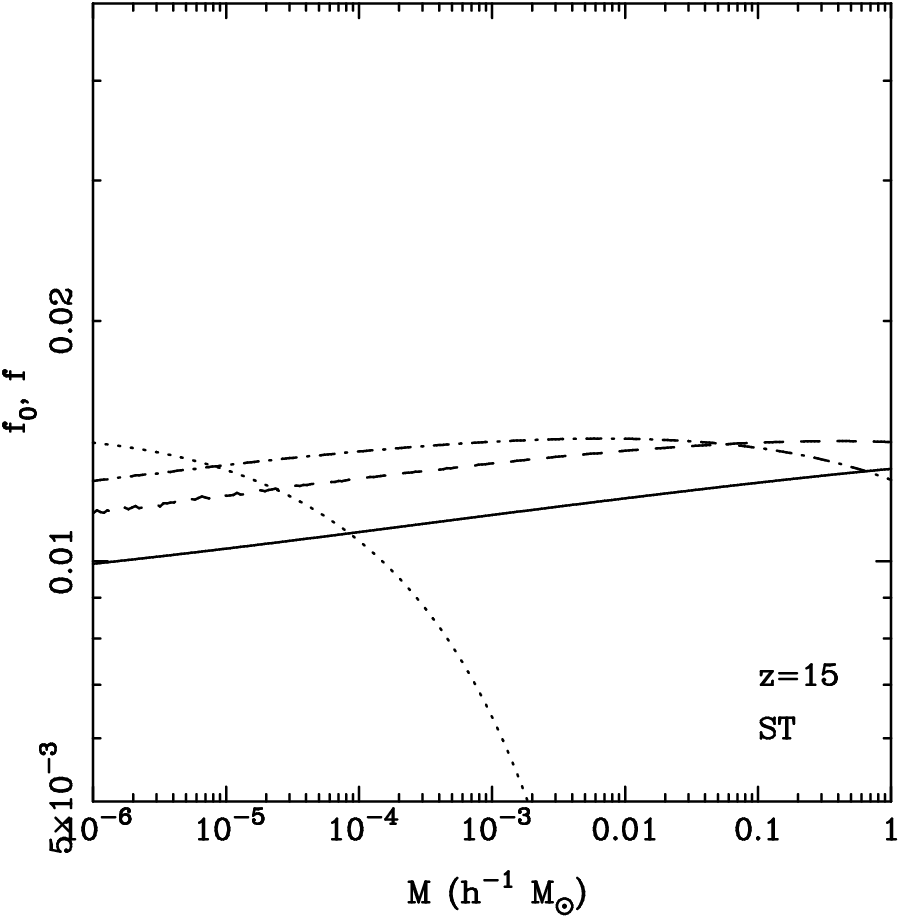}
\end{minipage}
\\
\end{tabular}
\caption{The multiplicity function expected in the $\Lambda$CDM model (see
  text for details).  The top row is for Press-Schechter mass function and the
  lower row is for Sheth-Tormen mass function.  The left column is for $z=20$
  and the right column is for $z=15$.  The expected multiplicity function is
  plotted as a function of mass, shown in each panel by a solid curve.  Other
  curves correspond to multiplicity for a finite simulation box: $\lbx =
  5$h$^{-1}$kpc (dotted curve), $\lbx = 20$h$^{-1}$kpc (dot-dashed curve) and
  $\lbx = 100$h$^{-1}$kpc (dashed curve).  These correspond to $C_1/\sigma_0^2
  \simeq 0.6$, $0.3$ and $0.19$, respectively.}
\end{figure*}

\subsubsection{N-Body Simulations}

We present here some preliminary results of a comparison of our
results with N-Body simulations.  
We do not try to fit either the Press-Schechter or the Sheth-Tormen mass
functions to simulations here, instead we use a simulation with a larger
$\lbx$ as reference and compare the number density of haloes as a function of
mass with another simulation run using a smaller $\lbx$. 
More detailed results obtained from N-Body simulations will be presented
in a later publication.  

We simulated the $n=-2$ power law model in an Einstein-de Sitter universe with
the normalisation $r_{nl}=8$Mpc at the final epoch. 
We chose one grid length of the simulation to equal $1$Mpc. 
The simulation was run with two values of the box-size: $\lbx = 64$Mpc and
$\lbx = 256$Mpc, with the latter being the reference. 
The simulations were run using the TreePM method
\citep{2002JApA...23..185B,2003NewA....8..665B}.  
The parallel TreePM code was used for the $256^3$ simulation
\citep{2004astro.ph..5220Ra}. 

Figure~5 shows the number density of haloes $n(M) dM$ in the mass range
$M~-~M+dM$ for these two simulations. 
Note that following our definitions $n(M) = f(M)/M^2$, where $f(M)$ is the
multiplicity function.
The solid line shows the number density of haloes in the reference
simulation.  
One can see the approximately power law variation at small $M$ and a rapid
fall off at large $M$.
Number density of haloes in the simulation with $\lbx = 64$Mpc is shown by the
dashed line. 
As expected from the above discussion, the deviation from power law starts at
smaller masses as the number density of very massive haloes is under-estimated
as compared to the reference simulation. 
At smaller $M$, we find about $10\%$ more haloes in this simulation as
compared to the reference. 
It is noteworthy that the number density of low mass haloes remains above that
in the reference simulation at all masses below the rapid drop around
$10^2$~M$_\odot$. 
Both the features follow the predictions in the preceeding discussion, indeed
we have shown that these features are independent of the specific analytical
form for the mass function. 
Here we have also shown that the same behaviour is reproduced in N-Body
simulations.  
A more detailed comparison is beyond the scope of this paper and the results
will be presented in a later publication.  


\section{Discussion}

In the preceeding sections, we have described a method to estimate errors in
the descriptors of clustering in the linear regime.  
We have also shown that the key results of the analytical study are borne out
by N-Body simulations.  
We have shown that the error is typically small if the scale of interest is
sufficiently smaller than the box size.  
An implicit requirement is that the scale of non-linearity too should be much
smaller than the box size; if this restriction is overlooked then we not
only ignore power in modes larger than the simulation box but also the
significant effects of mode coupling from scales in the mildly non-linear
regime. 
Therefore, we require $r, ~~ r_{nl} \ll \lbx$.  

We propose using $\sigma_1^2/\sigma_0^2$ as an indicator of the significance
of the finite box size, any descriptor of second moment can be used but
$\sigma$ has the virtue of being positive definite at all scales. 
Our proposal is that $\sigma_1^2(r)/\sigma_0^2(r), ~~
\sigma_1^2(r_{nl}(z))/\sigma_0^2(r_{nl}(z)) \ll 1$, for the finite box-size
corrections to be ignorable.  
All the $\sigma$s are evolved linearly here.
Conversely, the ratio $\sigma_1/\sigma_0$ at the scale of interest is
indicative of the magnitude of correction due to the finite box-size. 
For a given relative magnitude of the correction term ($\sigma_1/\sigma_0$),
$\lbx/r_{nl}$ is required to be larger for spectra with more large scale
power.  
Indeed, the required $\lbx/r_{nl}$ approaches infinity as the slope of the
power spectrum approaches $-3$.  

As a result of finite box-size corrections, the amplitude of density
perturbations is not a power law and the range of scales over which it can be
approximated by one becomes smaller as we approach $n+3 \rightarrow 0$. 
In the linear regime, the radial pair velocity is related directly with
$\bar\xi$ \citep{1980lssu.book.....P,1994MNRAS.271..976N}. 
As $\bar\xi$ is not a pure power law in simulations due to box-size
corrections, we expect that the pair velocities must also deviate from
expected values.  
This, in turn leads to deviations from scale invariant growth of density
perturbations. 
This explains the difficulty in getting scale invariant evolution for models
like $n=-2$ in N-Body simulations
\citep{1996ApJ...456...43J,1998ApJ...509..517J}. 
For realistic models like the $\Lambda$CDM, the correction term is significant
only if the scales of interest are below a few kpc and becomes larger as we
move to smaller scales (see Figure~2).  
Indeed, at these small scales we may require $\lbx/r \sim 10^4$ or even
greater in order to manage $C_1/\sigma_0^2 = 0.1$. 
Of course, a bigger simulation volume is required if we demand better
accuracy. 
On the other hand, if we are interested in scales larger than
$10^2$~kpc, present day simulations are sufficient for keeping $C_1/\sigma_0^2
\leq 0.01$. 

We have shown that at sufficiently small scales the correction due to a finite
box size can be written as a series of progressively smaller terms. 
The first correction term ($C_1$) is shown to be positive definite.  
We have also shown that the first correction term is the same for two point 
correlation function and $\sigma^2$, indeed it is the same for all descriptors
of the second moment for which the effective window function goes to unity at
small $k$. 

As an application of our method, we discussed the correction to mass function
and multiplicity function using the Press-Schechter as well as the
Sheth-Tormen approach.  
We have given the explicit form of the correction term due to finite box size
in each case.  
We have also given approximate expressions for the correction term and have
shown that the approximation is very good in case of mass function. 
The mass function is always under estimated in simulations due to finite
box-size corrections.  
Multiplicity function, and hence also the number density of haloes of a given
mass are underestimated at $M > M_{nl}$. 
At smaller mass scales, however, the multiplicity function is over estimated
and we find more haloes in a simulation than expected in the model.  
The mass scale at which the cross over from under estimate to over estimate
occurs is given by Eqn.(10) for Press-Schechter and Eqn.(14) for Sheth-Tormen
mass function.  

The over estimate at small scales is related to the under estimate of mass in
haloes at large scales.  
If the full power spectrum had been taken into account, the smaller haloes
would have merged to form more massive haloes. 
In absence of large scale modes, the formation of massive haloes is slowed
down and a larger number of low mass haloes survive. 
A detailed analysis of the effect of finite box size correction on merger
rates for haloes will be presented in a forthcoming paper.  
Of significant interest is the impact on rates of major mergers
\citep{2001MNRAS.325.1053C} as these have implications for observations.

We find that the over estimate in multiplicity function is large whenever the
ratio $\sigma_1^2(r,\lbx)/\sigma_0^2(r) \sim C_1(\lbx)/\sigma_0^2(r)$ is
large.  
To illustrate this correlation, we have plotted the multiplicity function
$f_0(M)$ for the $\Lambda$CDM model in Figure~6. 
This has been plotted for redshift $z=20$ and $z=15$ and the mass range has
been chosen such that very large box size is required to keep
$\sigma_1^2(r,\lbx)/\sigma_0^2(r)$ smaller than $0.1$.  
We have also plotted $f(M,\lbx)$ here, with $\lbx = 5$h$^{-1}$kpc (dotted
curve), $\lbx = 20$h$^{-1}$kpc (dot-dashed curve) and $\lbx = 100$h$^{-1}$kpc
(dashed curve).  
These correspond to $C_1/\sigma_0^2 \simeq 0.6$, $0.3$ and $0.19$,
respectively.  
The top row is for the Press-Schechter mass function and the lower row is
for the Sheth-Tormen mass function. 
An identical $x-y$ range has been used to highlight the differences between
the two models for mass function as well.  
It is noteworthy that the relative error is similar in both the cases even
though the multiplicity function itself is different. 
At $z = 20$, the multiplicity function is under estimated by a large amount
for $\lbx = 5$h$^{-1}$kpc, even though $\lbx/r_{nl} \simeq 120$ and if we are
interested in scales around $1$~pc then $\lbx/r \simeq 5000$.  
The situation at small masses is better for the other two simulation volumes
considered here.  
For $z = 15$, the scale of non-linearity is $r_{nl} = 1.4$h$^{-1}$kpc, very
close to $\lbx = 5$h$^{-1}$kpc and hence we do not expect believable results
for this box-size. 
Indeed, the two panels on the right demonstrate the large errors and the
absurdly incorrect shape of the multiplicity function. 
The difference in $f(M)$ and $f_0(M)$ at $10^{-6}M_\odot$ is about $25\%$ for
$C_1(\lbx)/\sigma_0^2=0.3$, in this case $\lbx=20$h$^{-1}$kpc and $\lbx/r
\simeq 2 \times 10^4$.  
The error in the multiplicity function is slightly larger than $10\%$ for
$\lbx=100$h$^{-1}$kpc even though $\lbx/r \simeq 10^5$ and $\lbx/r_{nl} \simeq
67$. 
The multiplicity function plotted here is the global function, and the
conditional mass functions should be used in order to estimate errors in
simulations where a high peak is studied at better resolution. 
Similar results are obtained for other mass functions that have been suggested
as a better fit to simulation data
\citep{2001MNRAS.321..372J,2005astro.ph..6395W}. 

The above discussion demonstrates the perils of using simulations where
$C_1(\lbx)/\sigma_0^2(r)$ is close to unity.   
One may argue that models for mass function have not been tested in this
regime where the local slope of the power spectrum is very close to $-3$, but
the fact that error in amplitude of density perturbations itself is large
should be reason enough to worry about reliability of results. 
Further, the agreement in the magnitude of errors for the several approaches
to mass functions also gives us some confidence in results. 

Majority of simulations are not affected by such serious errors, as the slope
of power spectrum approaches $-3$ only at very small scales (large wave
numbers). 
However, high resolution simulations of earliest structure formation in the
$\Lambda$CDM model need to have a very large dynamic range before the results
can be believed within $10\%$ of the quoted value. 
Indeed, our work may have some relevance to the ongoing discussion about the
Earth mass haloes
\citep{2005Natur.433..389D,2005astro.ph..2049Z,2005astro.ph..2118Z,2005astro.ph..2213M,2005astro.ph..8215Z}. 

\section{Conclusions}

Conclusions of this work may be summarised as follows.
\begin{itemize}
\item
We have presented a formalism that can be used to estimate the deviations of
cosmological N-Body simulations from the models being simulated due to the use
of a finite box size.
These deviations/errors are independent of the specific method used for doing
simulations. 
\item
For a given model, the deviations can be expressed as a function of the scale
$r$ of interest and $\lbx$, the box size of simulations.
\item
We have applied the formalism to study deviations in {\it rms}\/ fluctuations
in mass in the initial conditions.
\item
We find that the errors are small except for models where the slope of the
power spectrum is close to $-3$ at scales of interest. 
\item
The errors in case of the $\Lambda$CDM model are significant if the scale of
interest is smaller than a kpc even if simulations as large as the Millennium
simulation \citep{2005Natur.435..629S} are used.
\item
We have studied errors in mass function in the Press-Schechter model, as well
as other models.  
\item
The main error due to a finite box size is that the number of high mass haloes
is under estimated. 
\item
The number of low mass haloes is over predicted in simulations if the box size
effects are important.  
This happens as low mass haloes do not merge to form the (missing) high mass
haloes. 
\item
We have verified these trends using N-Body simulations. 
\end{itemize}

We note that it is extremely important to understand the sources
of errors in N-Body simulations and the magnitude of errors in quantities of
physical interest.  
N-Body simulations are used to make predictions for a number of observational
projects and also serve as a test bed for methods. 
In this era of ``precision cosmology'', it will be tragic if simulations prove
to be a weak link. 
We would like to note that our results apply equally to all methods of doing
cosmological N-Body simulations, save those where techniques like MAP are used
to include the effects of scales larger than the simulation volume. 

The method for estimating errors due to a finite box-size described in this
paper can be used for several physical quantities.  
In this paper we have used the method to study errors in clustering properties
and mass functions.  
We are studying the effect of finite box size on velocity fields and related
quantities, the results will be presented in a later publication.


\section*{Acknowledgements}

Numerical experiments for this study were carried out at cluster computing
facility in the Harish-Chandra Research Institute
(http://cluster.mri.ernet.in).
This research has made use of NASA's Astrophysics Data System. 
We thank the anonymous referee for useful comments.


\label{lastpage}


\begin{thebibliography}{}

\bibitem[\protect\citeauthoryear{{Bagla}}{2002}]{2002JApA...23..185B} {Bagla}
  J.~S.,  2002, Journal of Astrophysics and Astronomy, 23, 185,
  astro-ph/9911025   

\bibitem[\protect\citeauthoryear{Bagla}{2005}]{2005CSci...88.1088B} Bagla
  J.~S., 2005, CSci, 88, 1088 

\bibitem[\protect\citeauthoryear{{Bagla} \&
    {Padmanabhan}}{1994}]{1994MNRAS.266..227B}  {Bagla} J.~S.,  {Padmanabhan}
    T.,  1994, MNRAS, 266, 227  

\bibitem[\protect\citeauthoryear{{Bagla} \&
    {Padmanabhan}}{1997a}]{1997MNRAS.286.1023B} {Bagla} J.~S.,  {Padmanabhan}
    T.,  1997a, MNRAS, 286, 1023

\bibitem[\protect\citeauthoryear{Bagla and
  Padmanabhan}{1997b}]{1997Prama..49..161B} Bagla J. S. and Padmanabhan
  T. 1997b, Pramana -- Journal of Physics 49, 161 

\bibitem[\protect\citeauthoryear{Bagla, Prasad \&
  Ray}{2005}]{2004astro.ph..8429B} Bagla J.~S., Prasad J., Ray S., 2005,
  MNRAS, 360, 194

\bibitem[\protect\citeauthoryear{{Bagla} \& {Ray}}{2003}]{2003NewA....8..665B}
  {Bagla} J.~S.,  {Ray} S.,  2003, New Astronomy, 8, 665 

\bibitem[\protect\citeauthoryear{Bagla \& Ray}{2005}]{2005MNRAS.358.1076B}
  Bagla J.~S., Ray S., 2005, MNRAS, 358, 1076  

\bibitem[\protect\citeauthoryear{Bardeen et al.}{1986}]{1986ApJ...304...15B}
  Bardeen J.~M., Bond J.~R., Kaiser N., Szalay A.~S., 1986, ApJ, 304, 15  

\bibitem[\protect\citeauthoryear{Barkana \& Loeb}{2004}]{2004ApJ...609..474B}
  Barkana R., Loeb A., 2004, ApJ, 609, 474 

\bibitem[\protect\citeauthoryear{{Bernardeau}
    et~al.}{2002}]{2002PhR...367....1B} {Bernardeau} F.,  {Colombi} S.,
    {Gazta{\~ n}aga} E.,  {Scoccimarro} R., 2002, Physics Reports, 367, 1

\bibitem[\protect\citeauthoryear{Bertschinger}{1998}]{1998ARA&A..36..599B}
Bertschinger E., 1998, ARA\&A, 36, 599

\bibitem[\protect\citeauthoryear{Bond et al.}{1991}]{1991ApJ...379..440B}
Bond J.~R., Cole S., Efstathiou G., Kaiser N., 1991, ApJ, 379, 440

\bibitem[\protect\citeauthoryear{{Brainerd}
  et~al.}{1993}]{1993ApJ...418..570B}
{Brainerd} T.~G.,  {Scherrer} R.~J., {Villumsen} J.~V.,  1993, ApJ, 418, 570

\bibitem[\protect\citeauthoryear{Cohn, Bagla and
    White}{2001}]{2001MNRAS.325.1053C} {Cohn}, J.~D. and {Bagla}, J.~S. and
    {White}, M., 2001, MNRAS, 325, 1053

\bibitem[\protect\citeauthoryear{{Cole}}{1997}]{1997MNRAS.286...38C}
{Cole} S.,  1997, MNRAS, 286, 38

\bibitem[\protect\citeauthoryear{{Colombi} et~al.}{1994}]{1994A&A...281..301C}
{Colombi} S.,  {Bouchet} F.~R.,   {Schaeffer} R.,  1994, A\&A, 281, 301

\bibitem[\protect\citeauthoryear{{Couchman} \&
    {Peebles}}{1998}]{1998ApJ...497..499C} {Couchman} H.~M.~P.,  {Peebles}
  P.~J.~E., 1998, ApJ, 497, 499 

\bibitem[\protect\citeauthoryear{{Davis} \&
  {Peebles}}{1977}]{1977ApJS...34..425D} {Davis} M.,  {Peebles} P.~J.~E.,
  1977, ApJS, 34, 425

\bibitem[\protect\citeauthoryear{Diemand, Moore, \&
  Stadel}{2005}]{2005Natur.433..389D} Diemand J., Moore B., Stadel J., 2005,
  Nature, 433, 389  
 
\bibitem[\protect\citeauthoryear{{Gelb} \&
  {Bertschinger}}{1994a}]{1994ApJ...436..467G} {Gelb} J.~M.,  {Bertschinger}
  E.,  1994a, ApJ, 436, 467 

\bibitem[\protect\citeauthoryear{{Gelb} \&
    {Bertschinger}}{1994b}]{1994ApJ...436..491G} {Gelb} J.~M.,  {Bertschinger}
    E.,  1994b, ApJ, 436, 491 

\bibitem[\protect\citeauthoryear{Gunn \& Gott}{1972}]{1972ApJ...176....1G} 
Gunn J.~E., Gott J.~R.~I., 1972, ApJ, 176, 1 
 
\bibitem[\protect\citeauthoryear{{Gurbatov} et~al.},
  {1989}]{1989MNRAS.236..385G} {Gurbatov} S.~N.,  {Saichev} A.~I.,
  {Shandarin} S.~F.,  1989, MNRAS, 236, 385

\bibitem[\protect\citeauthoryear{{Hamilton}
    et~al.}{1991}]{1991ApJ...374L...1H} {Hamilton} A.~J.~S.,  {Kumar} P.,
    {Lu} E.,    {Matthews} A.,  1991, ApJL, 374, L1

\bibitem[\protect\citeauthoryear{Hawkins et al.}{2003}]{2003MNRAS.346...78H}
  Hawkins E. et al. 2003, MNRAS 346, 78

\bibitem[\protect\citeauthoryear{{Hui} \&
    {Bertschinger}}{1996}]{1996ApJ...471....1H} {Hui} L.,  {Bertschinger} E.,
    1996, ApJ, 471, 1 

\bibitem[\protect\citeauthoryear{{Jain} et~al.}{1995}]{1995MNRAS.276L..25J}
  {Jain} B.,  {Mo} H.~J.,    {White} S.~D.~M.,  1995, MNRAS, 276, L25 

\bibitem[\protect\citeauthoryear{Jain \&
    Bertschinger}{1996}]{1996ApJ...456...43J} Jain B., Bertschinger E., 1996,
  ApJ, 456, 43  

\bibitem[\protect\citeauthoryear{Jain \&
    Bertschinger}{1998}]{1998ApJ...509..517J} Jain B., Bertschinger E., 1998,
  ApJ, 509, 517  

\bibitem[\protect\citeauthoryear{Jenkins et al.}{2001}]{2001MNRAS.321..372J}
  Jenkins A., Frenk C.~S., White S.~D.~M., Colberg J.~M., Cole S., Evrard
  A.~E., Couchman H.~M.~P., Yoshida N., 2001, MNRAS, 321, 372  
 
\bibitem[\protect\citeauthoryear{{Kanekar}}{2000}]{2000ApJ...531...17Ka}
  {Kanekar} N.,  2000, ApJ, 531, 17 

\bibitem[\protect\citeauthoryear{Kauffmann \&
    Melott}{1992}]{1992ApJ...393..415K} Kauffmann G., Melott A.~L., 1992, ApJ,
    393, 415 

\bibitem[\protect\citeauthoryear{{Little} et~al.}{1991}]{1991MNRAS.253..295L}
  {Little} B.,  {Weinberg} D.~H.,    {Park} C.,  1991, MNRAS, 253, 295 

\bibitem[\protect\citeauthoryear{{Ma}}{1998}]{1998ApJ...508L...5M} {Ma} C.,
  1998, ApJL, 508, L5 

\bibitem[\protect\citeauthoryear{{Matarrese}
  et~al.}{1992}]{1992MNRAS.259..437M} {Matarrese} S.,  {Lucchin} F.,
  {Moscardini} L.,    {Saez} D.,  1992, MNRAS, 259, 437

\bibitem[\protect\citeauthoryear{Moore et al.}{2005}]{2005astro.ph..2213M}
  Moore B., Diemand J., Stadel J., Quinn T., 2005, arXiv:astro-ph/0502213 

\bibitem[\protect\citeauthoryear{{Nityananda} \&
    {Padmanabhan}}{1994}]{1994MNRAS.271..976N} {Nityananda} R.,  {Padmanabhan}
    T.,  1994, MNRAS, 271, 976 

\bibitem[\protect\citeauthoryear{Padmanabhan}{1993}]{1993sfu..book.....P} 
Padmanabhan T., 1993, {Structure Formation in the Universe}. Cambridge
University Press.

\bibitem[\protect\citeauthoryear{{Padmanabhan}}{1996}]{1996MNRAS.278L..29P}
  {Padmanabhan} T.,  1996, MNRAS, 278, L29 

\bibitem[\protect\citeauthoryear{{Padmanabhan}}{2002}]{2002tagc.book.....P}
  {Padmanabhan} T.,  2002, {Theoretical Astrophysics, Volume III: Galaxies and
  Cosmology}. Cambridge University Press. 

\bibitem[\protect\citeauthoryear{{Padmanabhan}
    et~al.}{1996}]{1996ApJ...466..604P} {Padmanabhan} T.,  {Cen} R.,
    {Ostriker} J.~P.,    {Summers} F.~J.,  1996, ApJ, 466, 604

\bibitem[\protect\citeauthoryear{{Peacock}}{1999}]{1999coph.book.....P}
  {Peacock} J.~A.,  1999, Cosmological physics. Cambridge University Press. 

\bibitem[\protect\citeauthoryear{{Peacock} \&
  {Dodds}}{1994}]{1994MNRAS.267.1020P} {Peacock} J.~A.,  {Dodds} S.~J.,  1994,
  MNRAS, 267, 1020 

\bibitem[\protect\citeauthoryear{{Peacock} \&
    {Dodds}}{1996}]{1996MNRAS.280L..19P} {Peacock} J.~A.,  {Dodds} S.~J.,
    1996, MNRAS, 280, L19 

\bibitem[\protect\citeauthoryear{{Peebles}}{1974}]{1974A&A....32..391P}
  {Peebles} P.~J.~E.,  1974, A\&A, 32, 391 

\bibitem[\protect\citeauthoryear{{Peebles}}{1980}]{1980lssu.book.....P}
  {Peebles} P.~J.~E.,  1980, {The large-scale structure of the universe}. 
  Princeton University Press.

\bibitem[\protect\citeauthoryear{{Peebles}}{1985}]{1985ApJ...297..350P}
  {Peebles} P.~J.~E.,  1985, ApJ, 297, 350 

\bibitem[\protect\citeauthoryear{Pope et al.}{2004}]{2004ApJ...607..655P} Pope
  Adrian C. et al. 2004, ApJ 607, 655

\bibitem[\protect\citeauthoryear{Power \& Knebe}{2005}]{2005astro.ph.12281P}
  Power C., Knebe A., 2005, arXiv:astro-ph/0512281  
 
\bibitem[\protect\citeauthoryear{Press \&
  Schechter}{1974}]{1974ApJ...187..425P} Press W.~H., Schechter P., 1974, ApJ,
  187, 425 

\bibitem[\protect\citeauthoryear{{Ray} \&
    {Bagla}}{2004}]{2004astro.ph..5220Ra} {Ray} S.,  {Bagla} J.~S.,  2004,
    astro-ph/0405220 

\bibitem[\protect\citeauthoryear{{Sahni} \&
    {Coles}}{1995}]{1995PhR...262....1S} {Sahni} V.,  {Coles} P., 1995,
    Physics Reports, 262, 1 

\bibitem[\protect\citeauthoryear{Sheth \& Tormen}{1999}]{1999MNRAS.308..119S}
  Sheth R.~K., Tormen G., 1999, MNRAS, 308, 119  

\bibitem[\protect\citeauthoryear{Sheth, Mo, \&
    Tormen}{2001}]{2001MNRAS.323....1S} Sheth R.~K., Mo H.~J., Tormen G.,
  2001, MNRAS, 323, 1  
 
\bibitem[\protect\citeauthoryear{Sirko}{2005}]{2005ApJ...634..728S} Sirko E.,
  2005, ApJ, 634, 728  

\bibitem[\protect\citeauthoryear{{Smith} et~al.}{2003}]{2003MNRAS.341.1311S}
  {Smith} R.~E.,  {Peacock} J.~A.,  {Jenkins} A.,  {White} S.~D.~M.,  {Frenk}
  C.~S.,  {Pearce} F.~R.,  {Thomas} P.~A.,  {Efstathiou} G.,    {Couchman}
  H.~M.~P.,  2003, MNRAS, 341, 1311 

\bibitem[\protect\citeauthoryear{Spergel et al.}{2003}]{2003ApJS..148..175S}
  Spergel D. N. et al., 2003, ApJS 148, 175

\bibitem[\protect\citeauthoryear{Springel et al.}{2005}]{2005Natur.435..629S}
  Springel V., et al., 2005, Nature, 435, 629  

\bibitem[\protect\citeauthoryear{{Tormen} \&
    {Bertschinger}}{1996}]{1996ApJ...472...14T} {Tormen} G.,  {Bertschinger}
    E.,  1996, ApJ, 472, 14 

\bibitem[\protect\citeauthoryear{Warren et al.}{2005}]{2005astro.ph..6395W}
  Warren M.~S., Abazajian K., Holz D.~E., Teodoro L., 2005,
  arXiv:astro-ph/0506395 

\bibitem[\protect\citeauthoryear{{Zel'dovich}}{1970}]{1970A&A.....5...84Z}
  {Zel'dovich} Y.~B.,  1970, A\&A, 5, 84 

\bibitem[\protect\citeauthoryear{Zentner, Koushiappas, \&
    Kazantzidis}{2005}]{2005astro.ph..2118Z} Zentner A.~R., Koushiappas S.~M.,
  Kazantzidis S., 2005, arXiv:astro-ph/0502118  

\bibitem[\protect\citeauthoryear{Zhao et al.}{2005}]{2005astro.ph..2049Z} Zhao
  H., Taylor J.~E., Silk J., Hooper D., 2005, arXiv:astro-ph/0502049  
 
\bibitem[\protect\citeauthoryear{Zhao et al.}{2005}]{2005astro.ph..8215Z}
  Zhao H., Taylor J.~E., Silk J., Hooper D., 2005, arXiv:astro-ph/0508215  
 
\end{thebibliography}
\end{document}